\documentclass[aps.pra,onecolumn,showpacs,preprintnumbers,amsmath,amssymb]{revtex4}
\usepackage{mathrsfs}
\usepackage{graphicx}
\usepackage{amsmath}
\usepackage{ulem}
\usepackage{color}

\begin{document}

\title{Dynamics of mesoscopic qubit ensemble coupled to cavity: role of collective dark states}
\author{A. A. Zhukov$^{1,2}$, D. S. Shapiro$^{1,3}$, W. V. Pogosov$^{1,4}$, Yu. E. Lozovik$^{1,5,6}$}

\affiliation{$^1$N. L. Dukhov All-Russia Research Institute of Automatics, 127055 Moscow, Russia}
\affiliation{$^2$National Research Nuclear University (MEPhI), 115409 Moscow, Russia}
\affiliation{$^3$V. A. Kotel'nikov Institute of Radio Engineering and Electronics, Russian Academy of Sciences, 125009 Moscow, Russia}
\affiliation{$^4$Institute for Theoretical and Applied Electrodynamics, Russian Academy of
Sciences, 125412 Moscow, Russia}
\affiliation{$^5$Institute of Spectroscopy, Russian Academy of Sciences, 142190 Moscow region,
Troitsk, Russia}
\affiliation{$^6$Moscow Institute of Physics and Technology, 141700 Moscow region, Dolgoprudny, Russia}

\begin{abstract}
We consider dynamics of a disordered ensemble of qubits interacting with single mode photon field, which is described by exactly solvable inhomogeneous Dicke model. In particular, we concentrate on the crossover from few-qubit systems to the system of many qubits and analyze how collective behavior of coupled qubits-cavity system emerges despite of the broadening. We show that quantum interference effects survive in the mesoscopic regime -- dynamics of an entangled Bell state encoded into the qubit subsystem remains highly sensitive to the symmetry of the total wave function. Moreover, relaxation of these states is slowed down due to the formation of collective dark states weakly coupled to light. Dark states also significantly influence dynamics of excitations of photon subsystem by absorbing them into the qubit subsystem and releasing quasiperiodically in time. We argue that predicted phenomena can be useful in quantum technologies based on superconducting qubits. For instance, they provide tools to deeply probe both collective and quantum properties of such artificial macroscopic systems.
\end{abstract}


\author{}
\maketitle
\date{\today }

\section{Introduction}

Controllable manipulation by quantum states of spin-photon coupled systems attracts a considerable current interest, since ensembles of spins/atoms interacting with the quantized electromagnetic field are promising candidates for implementation of quantum information and computation devices \cite{1,2,3,Moiseev}. One of the most perspective applications of such systems is storage of quantum information \cite{mem1,Molmer,Molmer1}. There are various physical realizations of spin-photon coupled systems, which range from superconducting artificial atoms (qubits) coupled to microwave resonators to nitrogen-vacancy (NV) centers in diamond and include even hybrid circuits combining two or more physical systems \cite{Norihyb}. Effective parameters of state-of-art spin-photon systems can be very different as well as numbers of coherently interacting spins (qubits). For instance, a typical number of NV centers in the ensemble is macroscopically large, while the coupling of a single center to the cavity mode is very weak. In contrast, state-of-art superconducting quantum circuits are limited by tens of qubits, while interaction between a single qubit and the microwave radiation can be relatively strong \cite{Macha,Zagoskin,Shulga,Norihyb,Girvin}. Nevertheless, essentially any solid-state physical realization is characterized by the inhomogeneous broadening of the density of states, i.e., by the splitting between excitation frequencies of individual spins. Broadening is caused by fundamental mechanisms and therefore is poorly controlled: for example, an excitation frequency of superconducting flux qubits depends exponentially on Josephson energies of contacts embedded into their structure \cite{qubit}, which makes it highly sensitive to characteristics of nanometer-scale Josephson junctions. For NV-centers, inhomogeneous broadening is induced by background disorder \cite{Lukin}.

Disorder in spin excitation energies is usually considered as a negative phenomenon, which prevents quantum information processing by introducing a decoherence \cite{Krimer1,Krimer2,Krimer3}. In order to overcome this problem, various ideas for the spectral engineering of spin density profile \cite{eng1,eng2} or spectral hole burning \cite{Krimer2} have been proposed (mostly in the context of NV centers in diamond). However, inhomogeneous broadening can also play a positive role since it may be utilized for the construction of multi-mode quantum memories \cite{Molmer}. Theoretical description of such systems is usually provided within Tavis-Cummings (Dicke) model, while interpretations are developed in terms of so-called radiant and subradiant modes, which turn out to be coupled if broadening is present in the system, since in this case they do not match exact eigenstates of the Dicke Hamiltonian \cite{Krimer1,Krimer2,Krimer3}. In other words, the excitation stored initially in the radiant mode is finally absorbed by a bath of subradiant states. A smart idea aimed to circumvent this problem is based on the hole burning technique: spins within certain 'dangerous' energy intervals, which are predominantly responsible for the interaction with subradiant modes, are neutralized for some time period by an external pulse \cite{Krimer2}. As a result,  special light-matter quantum states can be engineered, these states being mostly localized within the spin subsystem. If energy dissipation in a cavity is much larger than in the spin subsystem, as usually valid for realizations based on natural quantum systems, by using this approach it is possible to efficiently protect quantum state from the cavity decay, which might lead to realizations of quantum memory prototypes.

The aim of the present paper is a general study of dynamics of inhomogeneously broadened spin ensembles of mesoscopic rather than macroscopic sizes. This is especially actual for possible realizations of such ensembles coupled to microwave resonators within superconducting platform (sometimes referred to as quantum metamaterials, see, e.g., Refs. \cite{Macha,Zagoskin,Shulga}) and perhaps some other future solid-state circuits. In particular, we focus on the crossover from the system of just few qubits to the macroscopic system and study how collective dynamical behavior emerges along this crossover. This is done using an exact solution of Dicke Hamiltonian via Bethe ansatz \cite{Gaudin,Lieb,Wiegmann}. We restrict ourselves to the regime of weak excitation, i.e., when there is no more that one excitation in the system. We also consider different initial conditions and show that they can result in qualitatively different dynamics. Our approach provides a simple and pictorial understanding for main features of system's evolution. It allows to obtain a direct access to Hamiltonian eigenstates, which can be classified as dark and bright, and their properties, as well as to study explicitly role of such states in the system's dynamics. In contrast to earlier studies \cite{Andreev,Yuzb,Loss1,Strongly}, we mostly concentrate on the dynamics starting from excitations within spin subsystem being motivated by recent experimental advances in hybrid systems \cite{Krimer1,Krimer2,Krimer3} and consider mesoscopic qubit ensembles.

We show that in the limit of just few spins there appear Rabi-like oscillations between spin excitations and photon mode irrespective of the initial condition (excitation either in  spin or photon subsystem), as expected. However, as spin number increases, dynamics becomes highly sensitive to initial conditions. The most counterintuitive behavior is found for the initial condition of excitation in the spin subsystem -- spin excited state becomes frozen through what we call Zeno-like effect \cite{Zeno}: its relaxation time grows with the number of spins, so that in the limit of infinite spin number the excited state does not decay at all. For finite systems, there appear periodic partial revivals of an excited state, while the period of revivals grows with the spin number. We also demonstrate that certain collective excitations encoded into the spin subsystem and characterized by a finite entanglement, such as an antisymmetric Bell state, become even more robust with respect to the influence of environment of remaining spins despite of the fact that entanglement is, in general, a very fragile entity. This collective state being constructed from a couple of spins, essentially does not decay at all even in presence of a bath of remaining spins, provided these two spins have excitation energies neighboring in the energy space. Nevertheless, for larger separation between spin excitation frequencies, the evolution remains highly sensitive to the symmetry of the wave function, i.e., to the minus or plus sign in the Bell state. This result highlights a nontrivial role of quantum interference effects for disordered ensembles in a mesoscopic regime. Since entanglement is a key resource for quantum technologies, while quantum interference effects are essential for the experimental demonstration of "quantumness" of artificial macroscopic systems, our conclusions might be important for applications. The Zeno-like effect we found is directly linked to the formation of a quasi-continuum of Hamiltonian eigenstates poorly coupled to light which we refer to as collective dark states. They have similarities with subradiant states of homogeneous model the latter states being totally uncoupled from light \cite{Cummings}. In the case of a single photon present in the system in the initial moment, dark eigenstates also significantly affect dynamics of a mesoscopic ensemble -- they absorb the photon into the spin subsystem and then periodically release it giving rise to sharp revivals.

Our results are potentially useful for quantum states protection, storage and engineering. We believe that they can be also used to deeply probe quantum mechanical nature as well as collective properties of mesoscopic ensembles of artificial macroscopic spins, such as superconducting qubits, coupled to cavities.


\section{Hamiltonian and preliminaries}

We consider an ensemble of two-level systems coupled to a single mode photon field. This coupled system is described by Dicke Hamiltonian of the form
\begin{eqnarray}
H=\sum_{j=1}^L \epsilon_j \sigma_{j}^+ \sigma_{j}^- + \omega a^{\dagger } a + g \sum_{j=1}^L (a^{\dagger }\sigma_{j}^- + a \sigma_{j}^+),
\label{Hamiltonian}
\end{eqnarray}
where $a^{\dagger }$ and $a$ correspond to the boson degree of freedom:
\begin{eqnarray}
[a, a^{\dagger }]=1,
\label{boson}
\end{eqnarray}
while $\sigma_{j}^\pm$, $\sigma_{j}^z$ correspond to the paulion degrees of freedom and describe a set of $L$ two-level systems:
\begin{eqnarray}
& [\sigma_{j}^+, \sigma_{j}^-]=2\sigma_{j}^z, \\&
[\sigma_{j}^z, \sigma_{j}^\pm] = \pm \sigma_{j}^\pm.
\label{paulion}
\end{eqnarray}

The Hamiltonian (\ref{Hamiltonian}) commutes with the operator of the total pseudo-particle number, i.e., the number of bosons plus the number of excited two-level systems. Let us denote this number as $M$. Pseudo-particles of Dicke model are often referred to as excitations (of noninteracting system), but they should not be confused with excited states within a sector of given $M$ (of interacting system). Namely, for any fixed $M$, there are different eigenstates of the Hamiltonian. At given $M$, the lowest energy state is the ground state, while others represent excited states. Note that ground state energies for different values of $M$ can be also quite different. For example, if interaction constant $g$ is large enough, a global ground state can be attained at some nonzero $M$. This behavior is closely related to the so called superradiant transition \cite{Garraway}.

Note that the Hamiltonian (\ref{Hamiltonian}) is based on rotating wave approximation, which neglects counterrotating terms of the form $g(a \sigma_{j}^- + a^{\dagger } \sigma_{j}^+)$. These terms do not conserve excitation number and it is known that they can be omitted provided the detuning between the cavity and spin is not too large, $|\epsilon_j - \omega| \ll \omega$, see, e.g., Ref. \cite{Lozovik}. However, counterotating processes have to be taken into account even in the resonance, but only in certain specific situations, such as parametric and periodic excitation of a coupled qubit-cavity system \cite{Shapiro}. In the situation we consider in this article, counterrotating terms can indeed be safely neglected, since we are mostly interested in the interaction between the spin ensemble and cavity, which are close to the resonance, and do not treat such parametric excitations.

We also introduce an operator $S^{\dagger }(\lambda)$ defined as
\begin{eqnarray}
S^{\dagger }(\lambda) = a^{\dagger }+\sum_{j=1}^{L}\frac{g}{\lambda-\epsilon_j}\sigma_{j}^+,
\label{Slambda}
\end{eqnarray}
which is parametrized by the energy-like quantity (rapidity) $\lambda$. The state of the form
\begin{eqnarray}
\prod _{n=1}^{M}S^{\dagger }(\lambda_n) |\downarrow \downarrow \ldots \downarrow, 0 \rangle
\label{state}
\end{eqnarray}
is an eigenstate of the Hamiltonian, provided rapidities satisfy a set of Bethe equations
\cite{Gaudin,Talalaev,Duk}
\begin{eqnarray}
\frac{\lambda_n-\omega}{g}+\sum_{m\neq n}^M \frac{2g}{\lambda_n - \lambda_m} - \sum_{j=1}^L \frac{g}{\lambda_n - \epsilon_j}=0,
\label{Bethe}
\end{eqnarray}
while the eigenenergies $E$ are expressed through the roots $\lambda_n$ as
\begin{eqnarray}
E = \sum_{n=1}^M \lambda_n.
\label{energybethe}
\end{eqnarray}
There are in general multiple solutions of Eq. (\ref{Bethe}), i.e., many sets $\{ \lambda_n \}$, which form an energy spectrum within a sector of a given $M$ according to Eq. (\ref{energybethe}).

We restrict ourselves to the regime of a single pseudo-particle, $M=1$. In this case, there is only single rapidity $\lambda$, which satisfies a single Bethe equation
\begin{eqnarray}
\frac{\lambda-\omega}{g} - \sum_{j=1}^L \frac{g}{\lambda - \epsilon_j}=0.
\label{Bethesingle}
\end{eqnarray}
This is a polynomial equation of order $L+1$ and has the same number of solutions, which we refer to as $\lambda^{(\alpha)}$. It can be readily extracted from Eq. (\ref{Bethesingle}) that all solutions are real. They correspond to different eigenstates of the Hamiltonian in the same sector $M=1$. The unnormalized eigenfunctions can be represented as
\begin{eqnarray}
|\Phi_{\alpha} \rangle= S^{\dagger }(\lambda^{(\alpha)}) |\downarrow \downarrow \ldots \downarrow, 0 \rangle.
\label{unnorm}
\end{eqnarray}
It is easy to find a norm as
\begin{eqnarray}
\langle \Phi_{\alpha}|\Phi_{\alpha} \rangle= 1+\sum_{j=1}^{L}\frac{g^2}{(\lambda^{(\alpha)}-\epsilon_j)^2}.
\label{norm}
\end{eqnarray}
The normalized wave function thus reads as
\begin{eqnarray}
|\varphi_{\alpha} \rangle = \frac{1}{\sqrt{\langle \Phi_{\alpha}|\Phi_{\alpha} \rangle}} |\Phi_{\alpha} \rangle.
\label{normalized}
\end{eqnarray}

In Appendix A, we show how these results can be used to analyze system's evolution starting from different initial conditions, but corresponding to the same sector $M=1$. This number is of course conserved during the free evolution.

Note that, in absence of inhomogeneous broadening, Bethe states of Dicke model do not form a complete set (see, e.g., Ref. \cite{Yudson}), since degenerate nonradiating states, decoupled from light, cannot be obtained through Eq. (\ref{Bethesingle}).

\section{Flat distribution with nearly constant density of states}

\subsection{Hamiltonian eigenstates}

In the limit of a single spin $L=1$, the Hamiltonian (\ref{Hamiltonian}) reduces to the well known Jaynes-Cummings Hamiltonian. In this case, there exist only two solutions of Bethe equation (\ref{Bethesingle}). These are shown schematically in Fig. 1, where a full resonance between spin excitation energy and photon frequency is assumed. In the case of many spins $L$, the set of Bethe roots becomes drastically different due to the splitting between spin energies. This situation is also illustrated schematically in Fig. 1 for distribution of spin energies having abrupt terminations. In this case, in addition to the two separated roots relevant for Jaynes-Cummings model, new roots do appear, which are confined between neighboring spin energies. Note that in the limit of strong interaction $g$, when splitting between spin energies is irrelevant, two separated roots become responsible for Rabi oscillations of the whole ensemble of spins with frequency $g\sqrt{L}$.

\begin{figure}[h]
\includegraphics[width=0.85\linewidth]{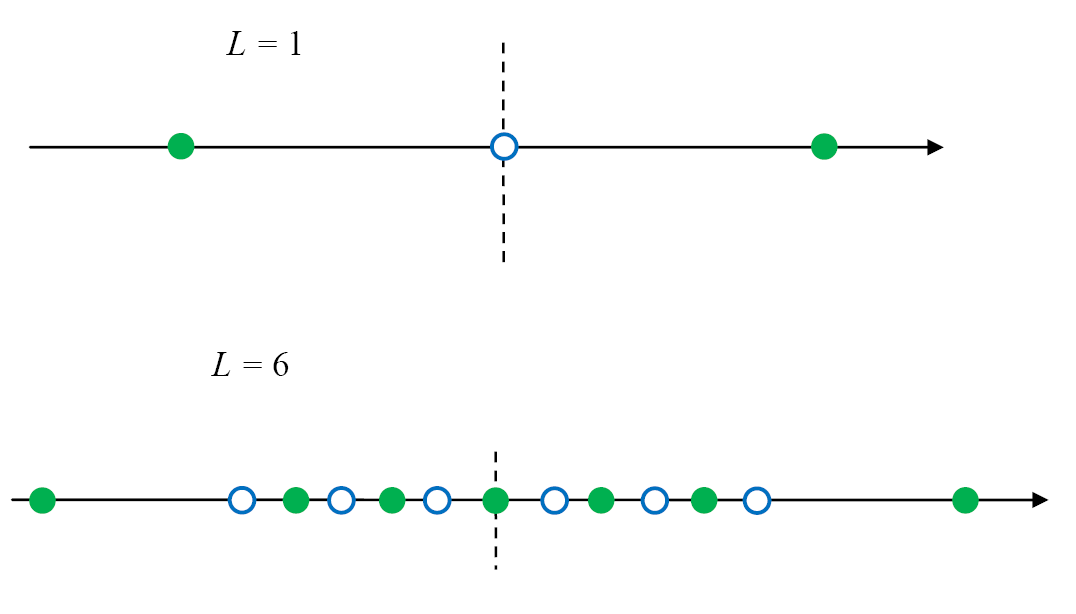}
\caption{
\label{cond}
Schematic illustration for the location of Hamiltonian eigenstates and spin excitation energies along the energy axis for different number of qubits $L=1$ and $L=6$. Blue open circles show positions of spin excitation energies. Green filled circles correspond to different Hamiltonian eigenenergies in the sector of single pseudoparticle, $M=1$. Vertical dashed line shows a position of photon energy, which is assumed to be in a resonance with mean spin excitation energy.}
\end{figure}

Among physically meaningful distributions of the density of states induced by disorder are the Gaussian and Lorentzian  distributions \cite{Littlewood} or $q$-Gaussian distribution relevant for NV-centers \cite{Krimer1}. Furthermore, simplified equally-spaced distribution of $\epsilon$ between the two cut-offs is of fundamental importance, since it allows to grasp main features of system's dynamics produced by splitting of excitation frequencies. Physically, it might correspond to the broad distribution, for which a central part, most strongly interacting with the photon mode, has nearly constant density of states. In the next Section, we will briefly describe the effects due to distributions with smooth tails.

Now let us concentrate on such an equally-spaced distribution of spin energies $\epsilon_j$ with the difference between neighboring $\epsilon_j $ denoted as $d$. We assume that the width of the distribution $\Omega$, i.e., the difference between maximum and minimum excitation energies, is independent on $L$, while the number of spins $L$ is large. We then consider the limit $L \rightarrow \infty$ and find explicitly a leading order in $L$ behavior as well as dominant corrections in $1/L$ essential for mesoscopic systems. In this limit, $d \rightarrow 0$. In order to construct $1/L$ expansion, we assume that $g\sqrt{L}$ is $L$-independent. Thus, $d/g \sim 1/\sqrt{L} \ll 1$ in this limit, so that there are many spin excitation frequencies within the energy scale $g$. We also focus on the most interesting regime, when $\Omega$ and $g\sqrt{L}$ are of the same order that results in a very rich phase diagram already for a static systems. While the former quantity is a natural scale to characterize broadening, the latter provides coupling energy between photon and spin subsystem in absence of broadening. Thus, at $\Omega \sim g\sqrt{L}$, there exists a pronounced competition between collective action of the whole ensemble of spins and their individual behavior  \cite{Littlewood,NucPhys2017}. The situation we consider is of particular importance in the context of superconducting qubits-resonator systems, since state-of-art superconducting circuits seem to start entering such a regime, where collective properties become significant despite of the disorder \cite{Macha,Zagoskin,Shulga,Norihyb}. Thus, we start from the limit of very few spins at $g \lesssim \Omega$, which is addressed mainly numerically, and analyze the whole crossover to the limit $g\sqrt{L} \gg \Omega$ with the particular focus on the intermediate mesoscopic regime $g\sqrt{L} \sim \Omega$.

In Appendix B, we address solutions to Bethe equation (\ref{Bethesingle}) within our model. There are in total $(L+1)$ solutions, $(L-1)$ of them are confined between neighboring spin excitation frequencies, while two remaining roots are, in general, separated, as shown in Fig. 1. Physically, two additional roots and confined roots describe Hamiltonian eigenstates with quite different properties. Indeed, it follows from Eq. (\ref{Slambda}) that each of these states $|\varphi^{(\alpha)}\rangle$ consists of a superposition of a single photon state and spin excited states with excitation energies detuned from $\lambda^{(\alpha)}$ by energy not too large, i.e., $ \lesssim g$. Therefore, eigenstates corresponding to confined roots are coupled essentially to each spin among a set of $\sim g/d \gg 1$ spins and to the photon mode. The coupling to the photon thus appears as quite weak. Therefore, such eigenstates can be characterized as dark states. Actually, they can be imagined as superpositions of many individual excited spins centered in energy space around a given spin, each superposition being only weakly coupled to light. In the limit of homogeneous model, these eigenstates should become fully decoupled from the light being gradually transformed to usual subradiant states. In contrast, two separated roots have smaller number of surrounding spin excitation energies, and coupling to the light for these two eigenstates is stronger, so that they can be refereed to as bright states. We would like to stress that collective dark states emerge only in the limit $g/d \gg 1$, since each of them must be represented by a superposition of many individual spin states in order to be dark. Indeed in the regime of just few spins and at $g \lesssim \Omega$ coupling to the light for all eigenstates is significant.

\subsection{Dynamics of the system with single spin excited in the initial moment}

Now we use our general results from Appendices A and B to study dynamics of the system with single spin excited in the initial moment, the excitation energy of this spin being $\epsilon_A$. We rewrite the time dependent wave function (\ref{explic}) in leading order as
\begin{eqnarray}
|\psi(t) \rangle \simeq \frac{d^2}{g^2\pi^2}\sum _{\alpha=1}^{L-1} e^{-i \lambda^{(\alpha)} t} \frac{g}{\lambda^{(\alpha)}-\epsilon_A} \frac{1}{1+\frac{1}{\pi^2}(\ln\frac{\epsilon_L-\epsilon_{\alpha}}{\epsilon_{\alpha}-\epsilon_{1}})^2}\left(a^{\dagger }+\sum_{j=1}^{L}\frac{g}{\lambda^{(\alpha)}-\epsilon_j}\sigma_{j}^+\right)|\downarrow \downarrow \ldots \downarrow, 0 \rangle.
\label{explic1}
\end{eqnarray}
Let us stress that we omitted in Eq. (\ref{explic1}) a contribution from two separated roots (bright states), which is justified in this order. For the amplitude of the probability to find the initially excited spin still in this state we have
\begin{eqnarray}
\langle \psi(t=0) |\psi(t) \rangle \simeq   \frac{d^2}{\pi^2}\sum _{\alpha=1}^{L-1} e^{-i \lambda^{(\alpha)} t} \frac{1}{(\lambda^{(\alpha)}-\epsilon_A)^2} \frac{1}{1+\frac{1}{\pi^2}(\ln\frac{\epsilon_L-\epsilon_{\alpha}}{\epsilon_{\alpha}-\epsilon_{1}})^2}.
\label{ampprob}
\end{eqnarray}

Let us now focus on the situation, when all energies $\epsilon_j$ are centered around $\omega$, while $\epsilon_A$ is also in a resonance with $\omega$. It is easy to see that, under such conditions, $\delta_{\alpha} \simeq d/2$ in a vicinity of $\omega$. It is also clear that the logarithm in the right-hand side of Eq. (\ref{ampprob}) can be omitted (at least for $t$ not too large). We then obtain in leading order for $t \lesssim 2\pi/d$
\begin{eqnarray}
\langle \psi(t=0) |\psi(t) \rangle \simeq   \frac{d^2}{\pi^2} e^{-i\epsilon_At} \sum _{\alpha=1}^{L-1}  e^{-i (\epsilon_{\alpha}-\epsilon_A) t} \frac{1}{(\epsilon_{\alpha}-\epsilon_A+d/2)^2}  \simeq \frac{8}{\pi^2} e^{-i\epsilon_At} \sum_{n=0}^{\infty}\frac{\cos\left[(2n+1)td/2\right]}{(2n+1)^2}.
\label{ampprob1}
\end{eqnarray}
This function represents a simple periodic triangle wave of a period $4\pi / d$. The initially excited spin decays on a time scale $\sim 1/d \sim L$. There is a revival after such a decay due to the finite size of the environment of other spins, i.e., the finiteness of the number of spins having frequencies close to $\omega$. In other words, an initial excitation becomes redistributed between $\sim g/d$ spins most strongly interacting with an initially excited spin via the photon field. After some time, there occurs a refocusing of such a collective state into an initial state. At long times, an ideal periodic function (\ref{ampprob1}) becomes somehow smeared and less regular, since various subdominant contributions in $1/L$ come into play, which, in particular, results in certain finite occupations of strongly detuned spins.

\begin{figure}[h]
\includegraphics[width=0.45\linewidth]{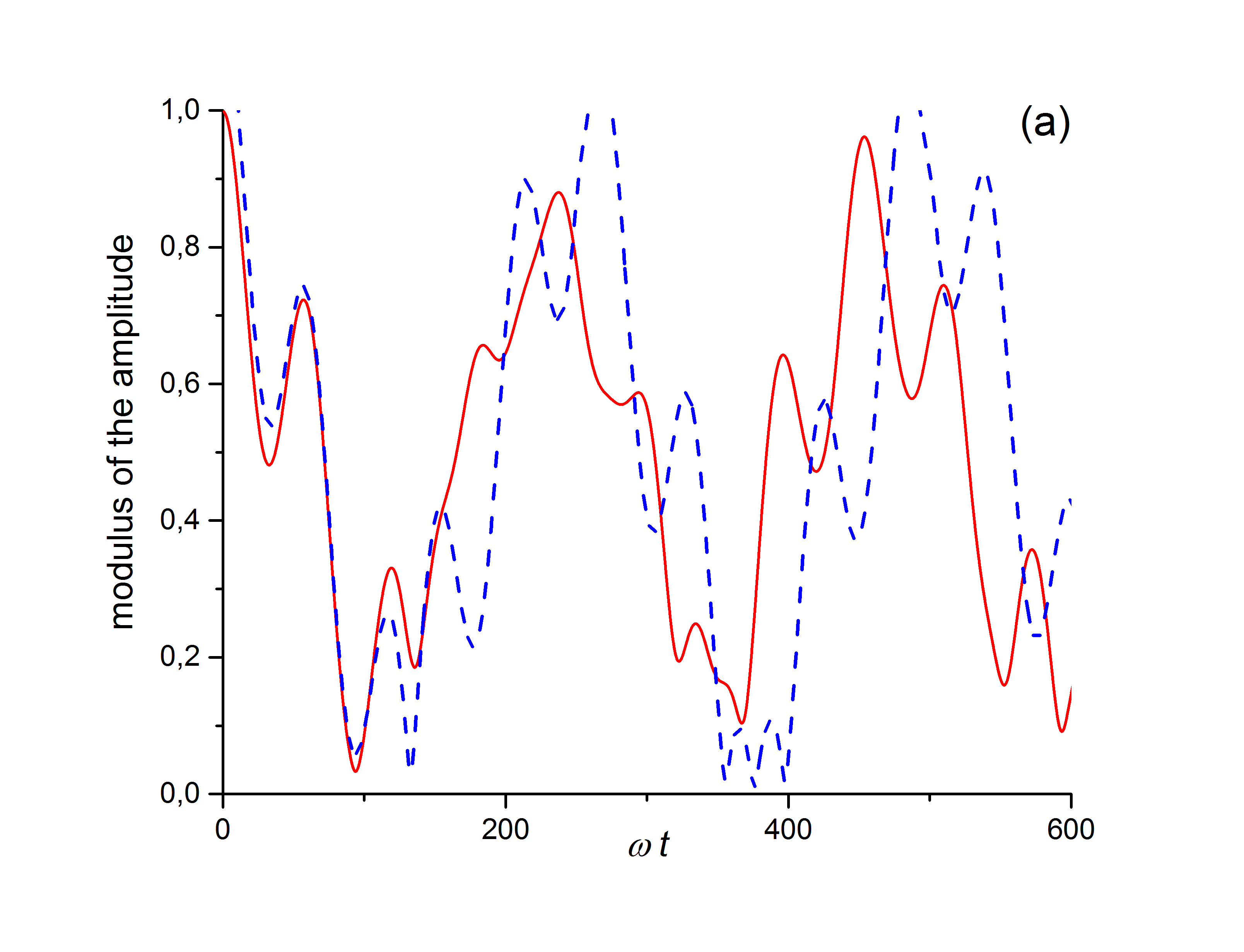}
\includegraphics[width=0.45\linewidth]{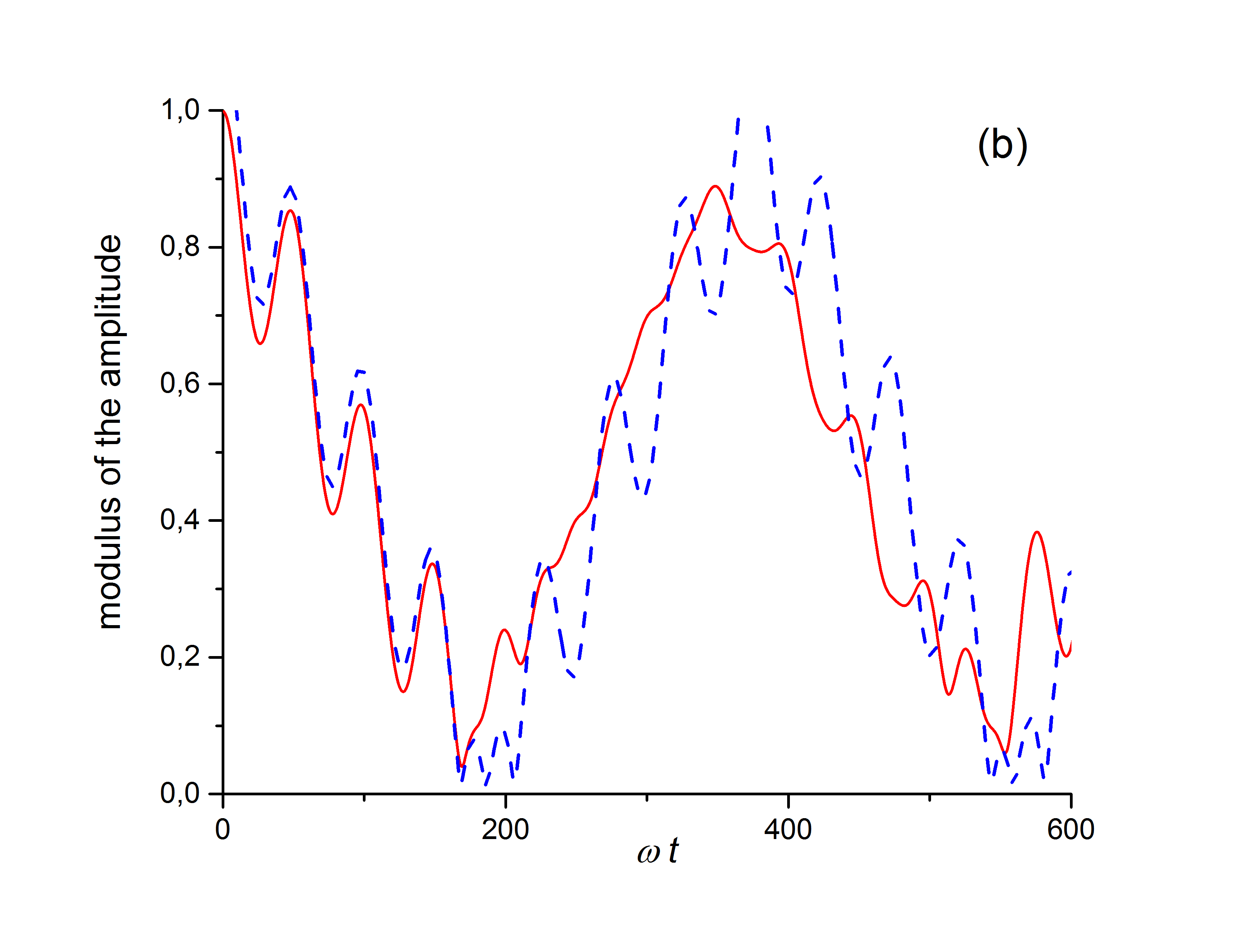}
\includegraphics[width=0.45\linewidth]{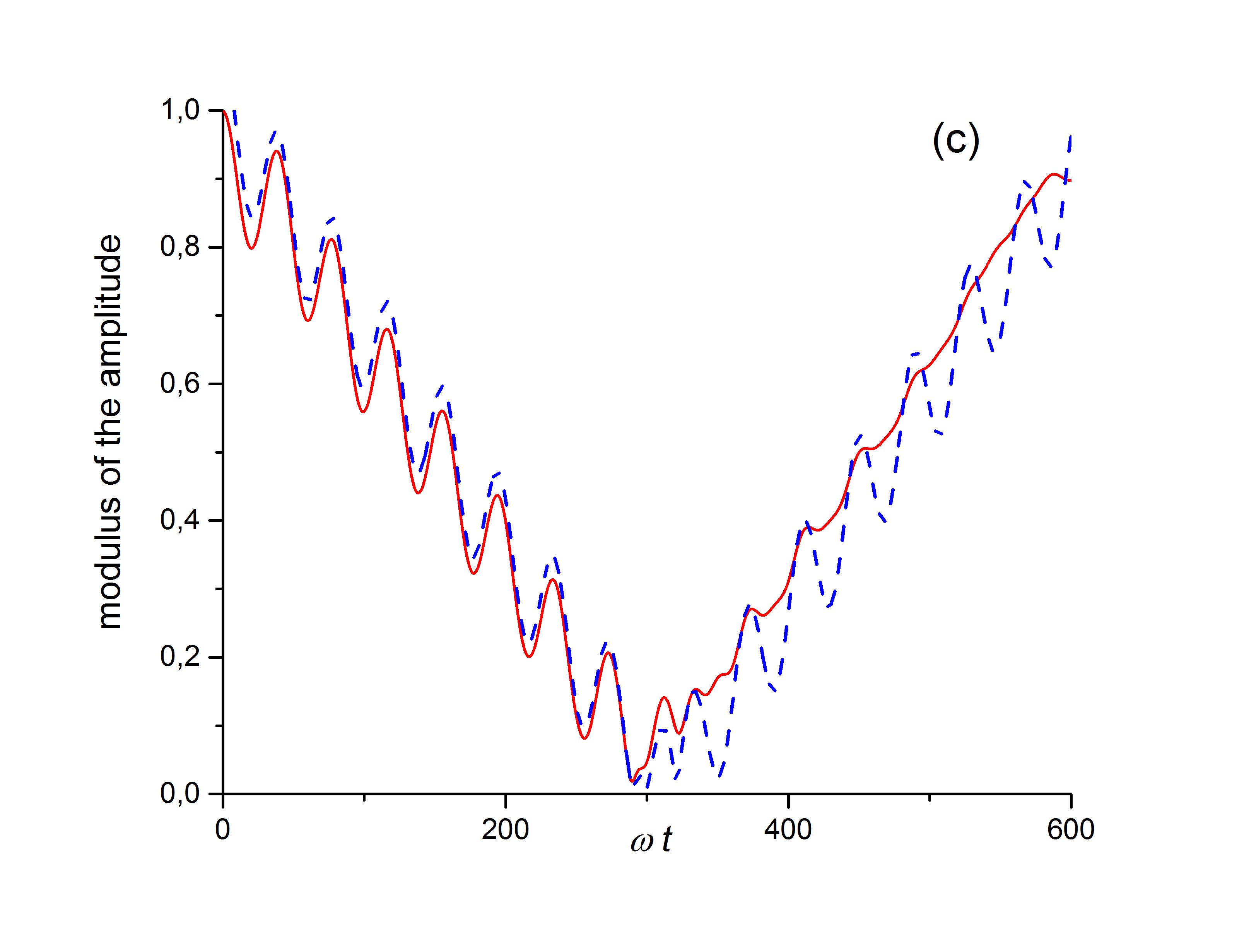}
\includegraphics[width=0.45\linewidth]{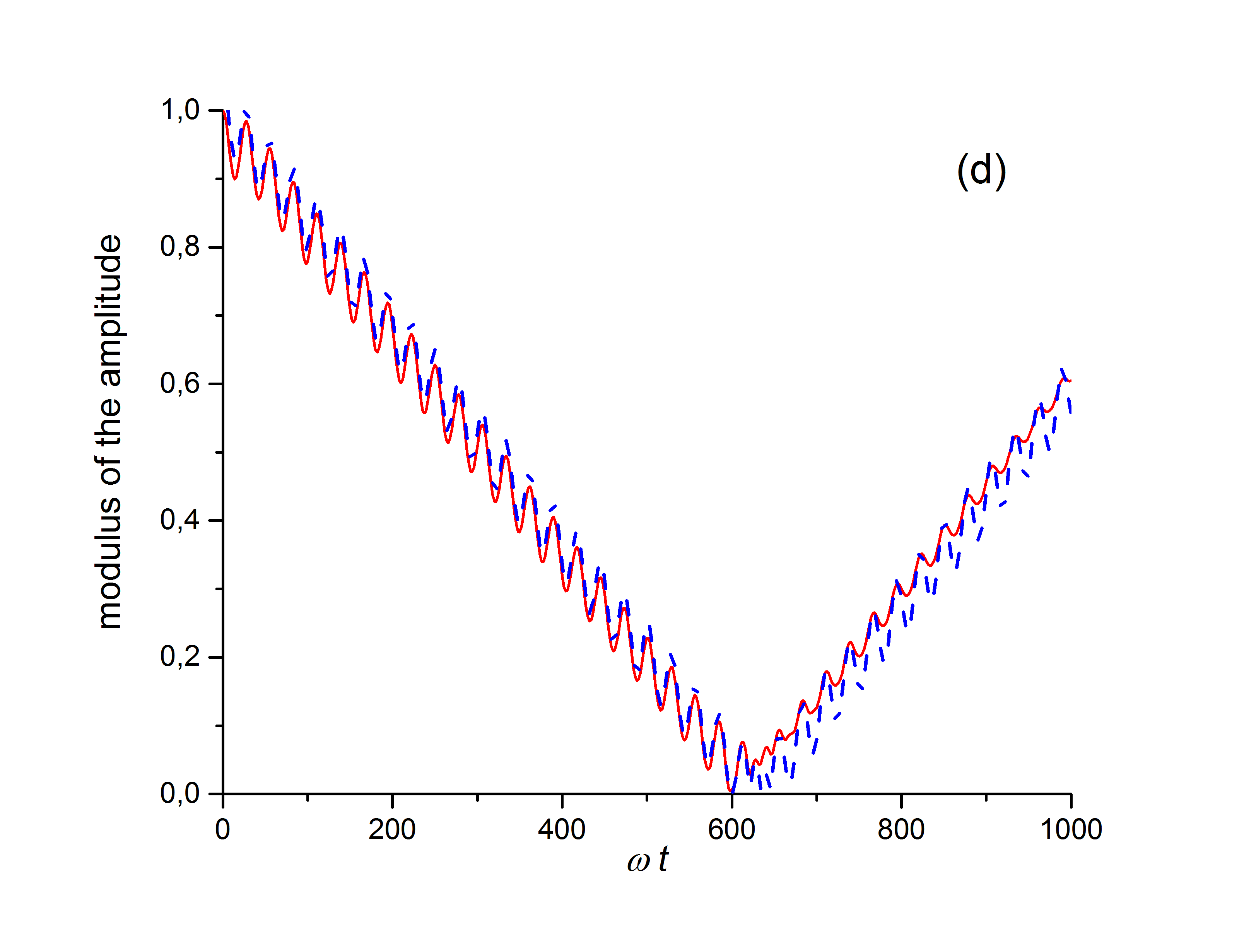}
\caption{
\label{spin}
Evolution of  $|\langle \psi(t=0) |\psi(t) \rangle|$ for $L=4$ (a), $L=6$ (b), $L=10$ (c), $L=20$ (d) and at $g=0.05\omega$, $\Omega = 0.1\omega$. The initial state of the system corresponds to the single spin excited among the ensemble of spins with broadened excitation energies. Red solid lines show numerical results, while blue dashed lines correspond to explicit results based on $1/L$ expansion.}
\end{figure}

The major conclusion deduced from Eq. (\ref{ampprob1}) is that the excited state decays on a time scale, which grows as the number of spins (density of states) in the system increases. Thus, the excited state becomes stabilized by the continuum of dark states giving rise to what we refer to as  Zeno-like effect. A similar effect of "radiation trapping" is known from literature for homogeneous Dicke model \cite{Cummings}. Thus, we show that such an effect also takes place for inhomogeneously broadened ensembles and we reveal how it emerges as the number of spins in such an ensemble grows.
Physically, this phenomenon might be attributed to the fact that the energy of the initial state with one of the spins excited is closer to the eigenenergies of dark states. The dynamics is then governed mostly by these dark states, which are weakly coupled to the light. Such an understanding has to be contrasted with naive expectations that an addition of extra 'parasitic' spins should only lead to a kind of a chaotization of Rabi oscillations between the given excited spin and the photon field. A chaotization occurs only in the limit of just few spins in the sense that in this limit dynamics is highly sensitive to precise detunings between spin excitation frequencies and the photon frequency. However, at $L \gg 1$ it transforms to the universal triangle wave dependence sensitive only to the mean density of states in the vicinity of $\omega$.

Let us now take into account a first order correction in $1/L$. The most important contribution in this order stems from the fact that we have neglected two additional roots of Eq. (\ref{Bethesingle}), i.e., the bright states. We can readily find that these states produce a correction to the amplitude of the probability (\ref{ampprob1}) given by
\begin{eqnarray}
\delta \langle \psi(t=0) |\psi(t) \rangle \simeq  \frac{1}{L} e^{-i\epsilon_At} \sum_{\alpha=0,L}  e^{-i t (\lambda^{(\alpha)}-\epsilon_A)} \frac{1}{\langle \Phi_{0,L}|\Phi_{0,L} \rangle} \frac{g^2L}{(\lambda^{(\alpha)}-\epsilon_A)^2}.
\label{ampprobcor}
\end{eqnarray}
Bright states give rise to Rabi-like oscillations contributing to the total dynamics.

Figure \ref{spin} visualizes how dynamics changes as $L$ grows. It shows the evolution of $|\langle \psi(t=0) |\psi(t) \rangle|$ for $L=4$ (a), $L=6$ (b), $L=10$ (c), $L=20$ (d) and at $g=0.05\omega$, $\Omega = 0.1\omega$, $\epsilon_A=\omega$. Two different results are compared -- the first one is obtained numerically by solving Bethe equation and the second one is an explicit result, which contains a dominant contribution (\ref{ampprob1}) as well as a subdominant correction (\ref{ampprobcor}). We indeed find quite good agreement between the numerical and explicit results starting from $L\approx10$, while qualitative agreement exists even for smaller values of $L$. These plots visualize how Rabi oscillations at $L=1$ transform into triangle wave of very large period at $L \gg 1$.

\begin{figure}[h]
\includegraphics[width=0.45\linewidth]{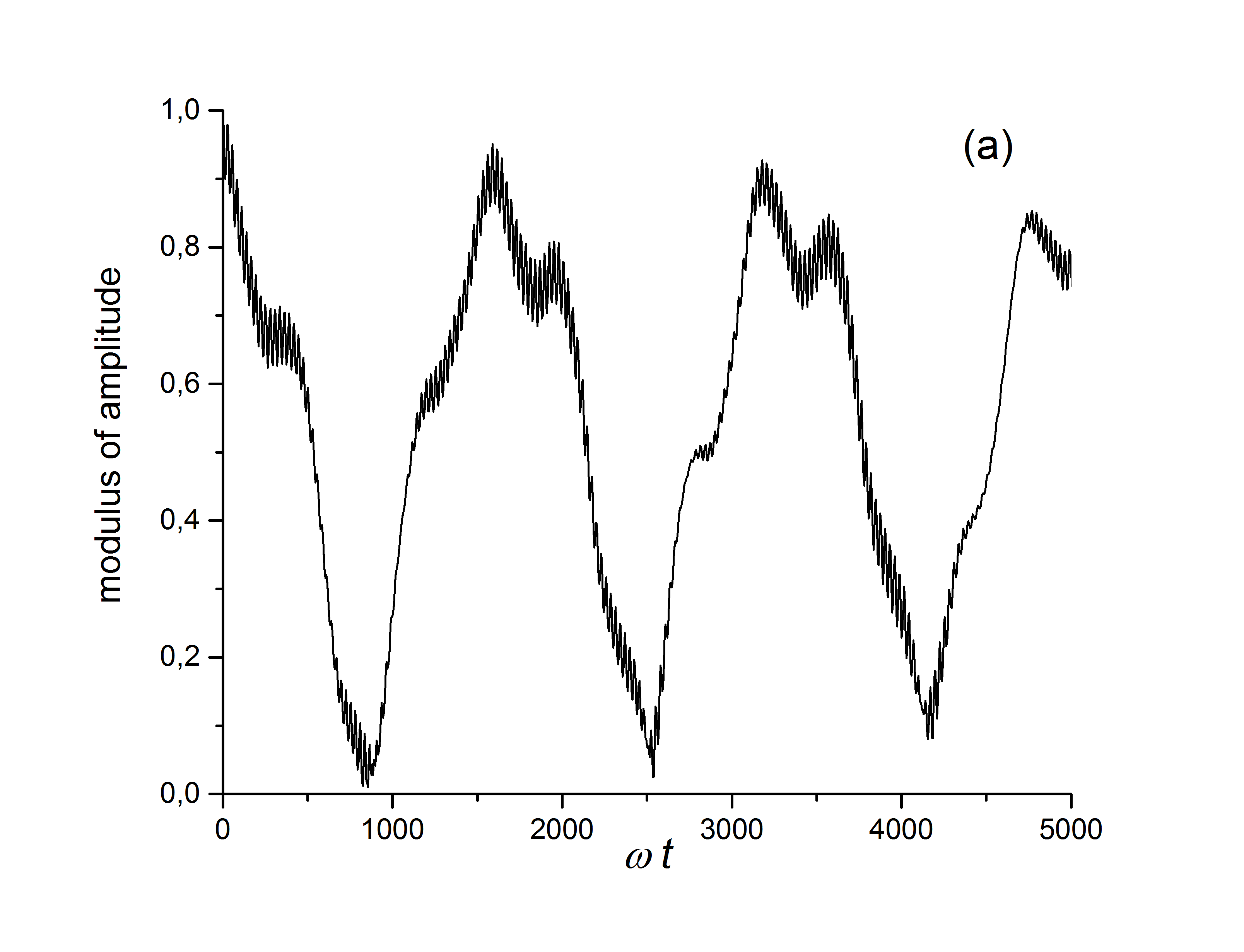}
\includegraphics[width=0.45\linewidth]{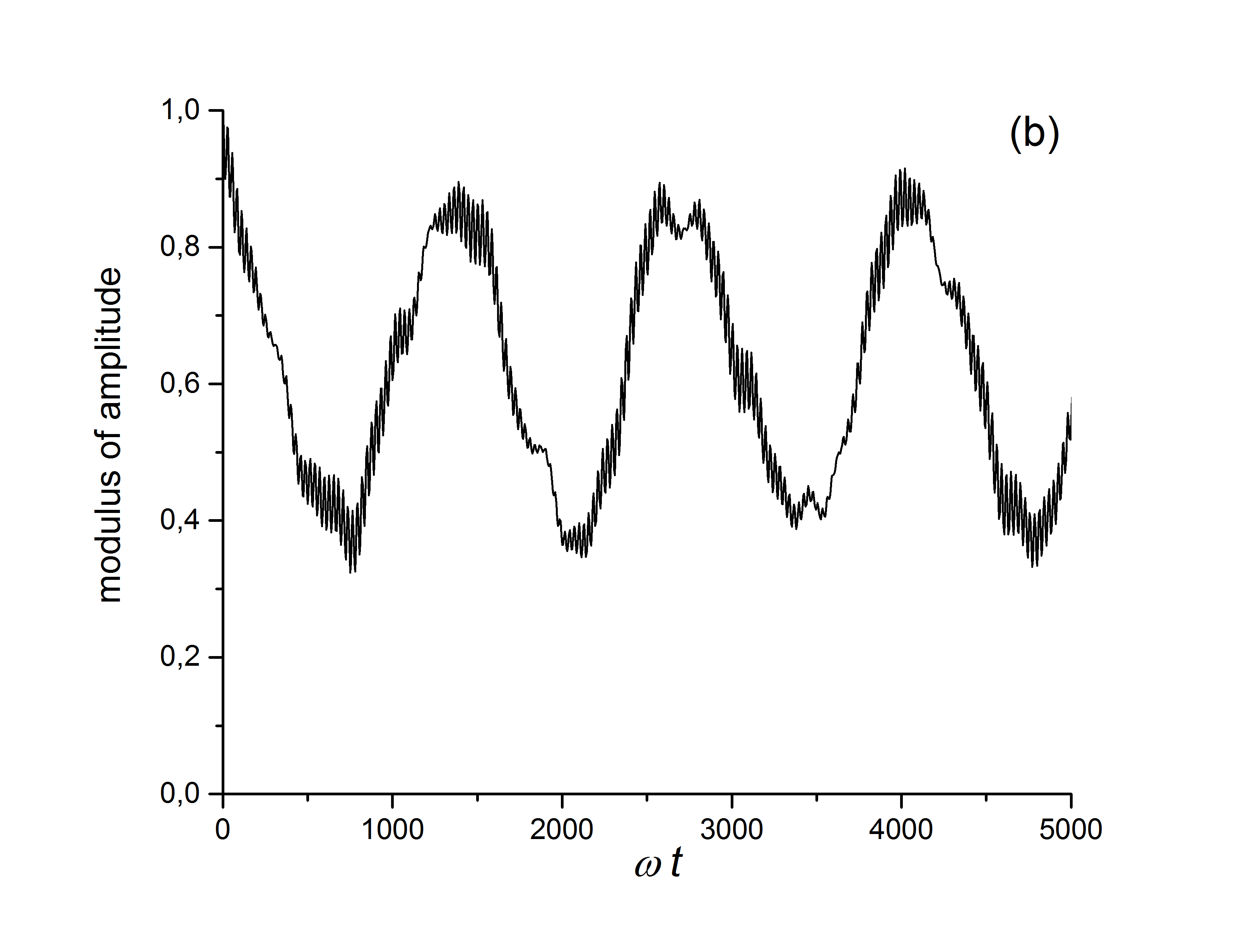}
\caption{
\label{rand}
Evolution of  $|\langle \psi(t=0) |\psi(t) \rangle|$ for $L=20$ (d) and at $g=0.05\omega$, $\Omega = 0.1\omega$ and two different realizations of disorder. Excitation energies of spins are randomly distributed between two cutoffs.}
\end{figure}

Despite of the universality in the large $L$ limit, actual time evolution for the spin occupation can be very sensitive to mesoscopic fluctuations. In order to illustrate this, we plot in Fig. \ref{rand} the same quantity $|\langle \psi(t=0) |\psi(t) \rangle|$ calculated numerically at $L=20$, $g=0.05\omega$, and $\Omega = 0.1\omega$, but for randomly distributed spin energies confined between the same cutoffs, while $\epsilon_A$ is taken to be closest to the resonance among all spins. Fig. \ref{rand} (a) and (b) correspond to two typical realizations of disorder. In the first case, the evolution of $|\langle \psi(t=0) |\psi(t) \rangle|$ is quite close to the similar dependence for the idealized equally-spaced distribution, while in the second case spin occupation oscillates with nearly the same period, but it does not reach zero. Such a behavior in the latter case is a direct consequence of mesoscopic fluctuations, i.e., finiteness of spin number $L$.

We also verify that Eq. (\ref{ampprob}) is consistent with the normalization condition $|\langle \psi(t=0) |\psi(t=0) \rangle|=1$. Indeed, it is known that $\sum_{n=0}^{\infty}\frac{1}{(2n+1)^2}=\pi^2/8$ this identity being connected to the famous Basel problem. We would like to stress that the way, this relation appeared in our formalism, seems to be highly nontrivial.

In contrast, occupations in the photon subsystem appear to be generally very small under the initial condition of single spin excited. Namely, the probability amplitude of finding a system in the state with a single photon vanishes in the limit $L \rightarrow \infty$. This result can be derived from Eq. (\ref{explic1}) yielding
\begin{eqnarray}
\langle 0, \downarrow \ldots \downarrow | a |\psi(t) \rangle \simeq \frac{4d}{g \pi^2}e^{-i\epsilon_At}
\sum_{n=0}^{\infty}\frac{\sin\left[(2n+1)td/2\right]}{2n+1},
\label{ampprobphot}
\end{eqnarray}
which is a periodical square wave of the amplitude proportional to $d/g$. It is small provided there are many spin excitation energies within $g$. Note that two separated roots produce an additional contribution of the order of $1/L$. This result is consistent with the fact that the coupling between the dark states and the photon state is very weak, while dark states are predominantly responsible for the system's dynamics in large $L$ limit. Thus, "radiation trapping" effect for homogeneous model \cite{Cummings} is naturally recovered. If energy dissipation is present in the system and it is much larger for the cavity compared to spin subsystem, than a coupling to dark states allows to drastically reduce total effective dissipation.

The obtained results are potentially important in the context of quantum information storage and quantum state protection, since they show that dark eigenstates are able to greatly enhance the life time of the single excited spin and also they reveal how stabilization of excitations within spin subsystem induced by these states emerges as the number of spins increases. Perhaps, our findings can be also used to probe properties of mesoscopic ensembles of artificial spins coupled to a cavity. By exciting one of the spins of the ensemble via an additional waveguide and tracing its evolution, it is possible to see whether the ensemble behaves collectively according to the scenario we predict or such a behavior is suppressed due to the disorder and/or decoherence processes.

\begin{figure}[h]
\includegraphics[width=0.45\linewidth]{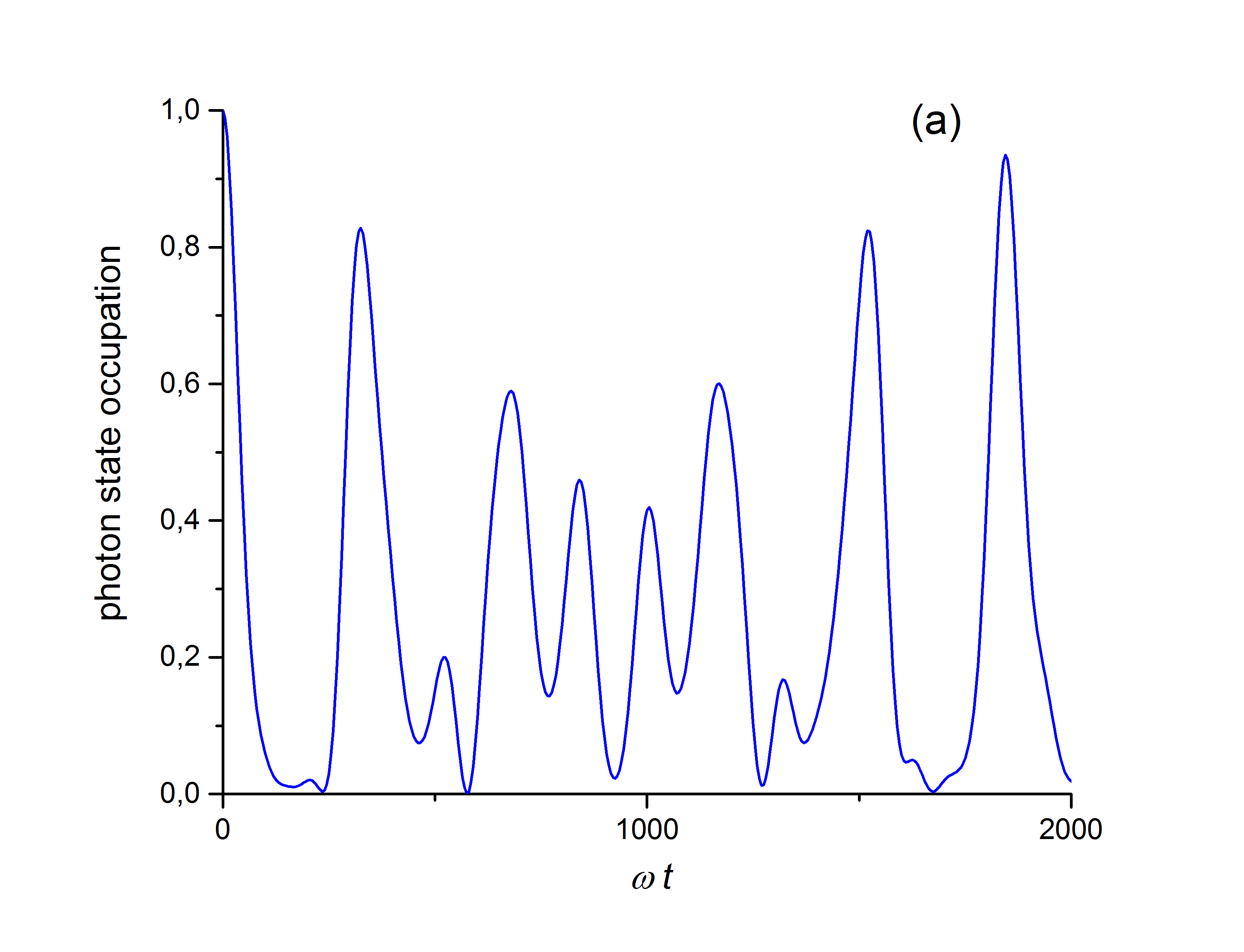}
\includegraphics[width=0.45\linewidth]{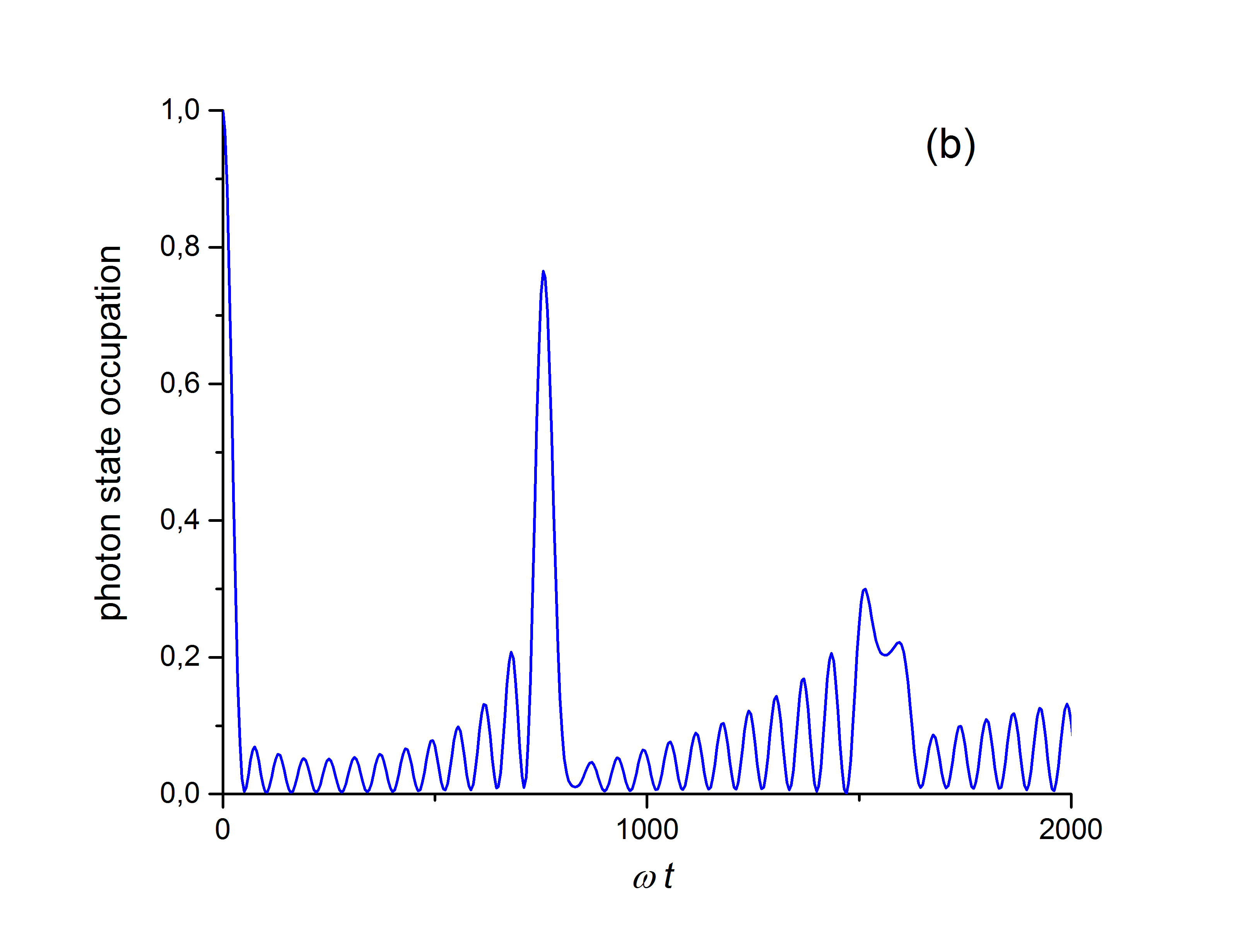}
\includegraphics[width=0.45\linewidth]{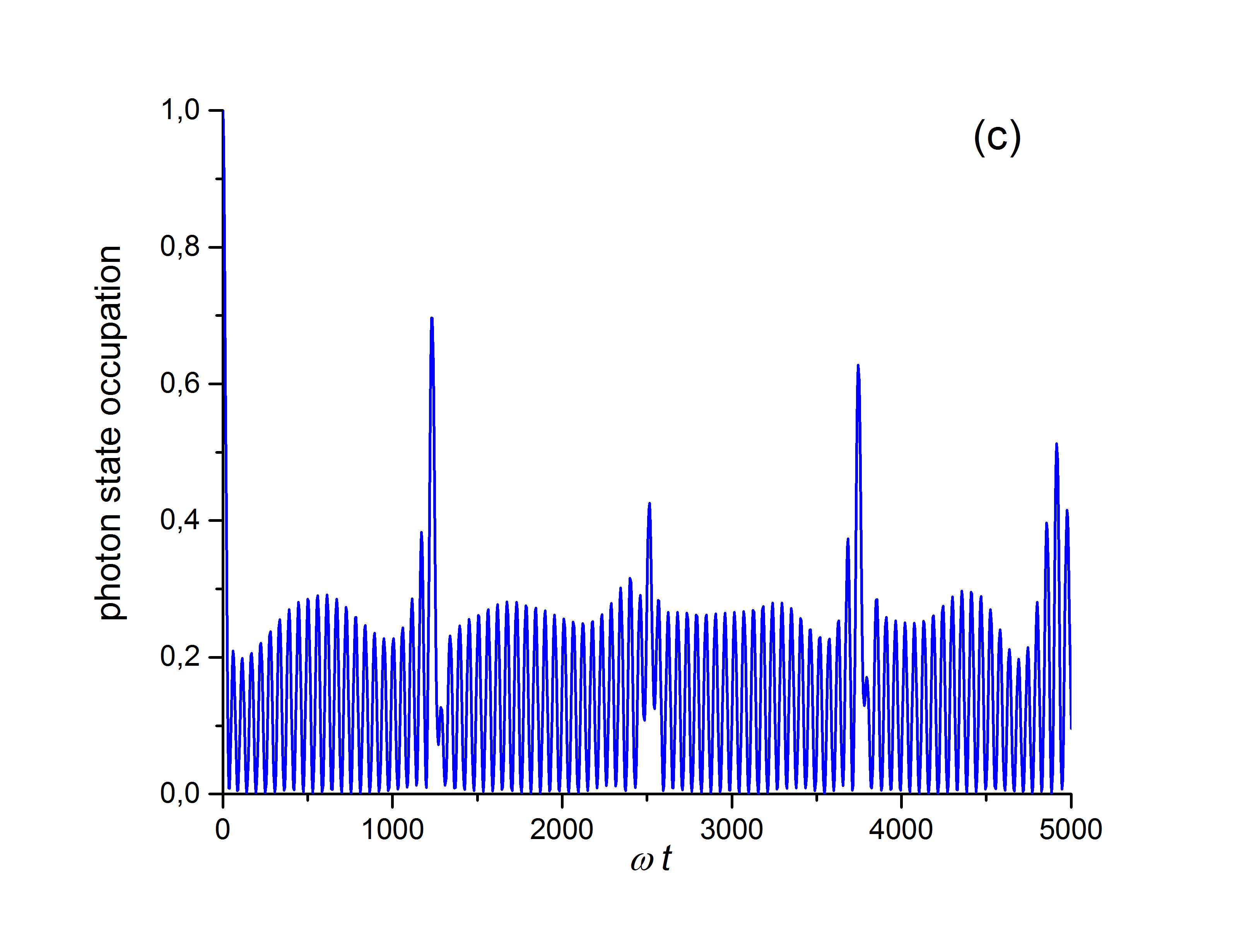}
\caption{
\label{photon}
Evolution of the occupation of a single photon state for $L=4$ (a), $L=8$ (b), $L=20$ (c) and at $g=0.05\omega$, $\Omega = 0.1\omega$}
\end{figure}

The problem we study might be also considered as an exactly solvable toy model for the coupled qubit-cavity system in presence of mesoscopic environment of 'parasitic' quantum two-level systems (fluctuators) interacting with the main system via photon degree of freedom and leading to decoherence. This model, however, is unable to account for the fluctuators own environment, so that it is applicable provided fluctuators are more strongly coupled to the qubit-cavity system than to their environment. The model has to be contrasted with the spin-boson model and it corresponds to pure quantum regime when an entanglement between the qubit-cavity system and fluctuators is of importance \cite{Galperin}. The existence of the exact solution allows to perform microscopic analysis for such a situation without switching to any simplified and not fully controllable approximation. In this context, we investigated the dynamics of a single qubit-cavity coupled system in presence of a mesoscopic ensemble of fluctuators with strongly detuned excitation frequencies. We found that such an ensemble produces very significant phase drift ('dephasing') for the amplitude of the probability of qubit excitation. We would like to stress that this effect is much stronger than the influence of fluctuators on qubit excited state population -- even if Rabi oscillations are almost perfectly reproduced, the phase drift can be significant. Of course, it becomes larger as the detuning between the mean fluctuator frequency and photon mode is decreased. We also expect that presence of individual environments of qubits should lead to the suppression of Zeno-like effect, so that in the regime of strong energy dissipation lifetime of the qubit excited state is limited by the decay into its own environment.

\subsection{Dynamics of the system with single photon state in the initial moment}

Let us now apply our general results for Bethe roots in the equally-spaced model to the dynamics of the system starting from a single photon state. For the amplitude of probability to have still a single photon state after some evolution, we obtain from (\ref{explicphot})
\begin{eqnarray}
\langle 0, \downarrow \ldots \downarrow | a |\psi(t) \rangle \simeq \frac{d^2}{\pi^2 g^2} e^{-i d t/2} \sum _{\alpha=1}^{L-1} e^{-i \epsilon_{\alpha} t} \frac{1}{1+\frac{1}{\pi^2}\left(\ln \frac{\epsilon_L-\epsilon_{\alpha}}{\epsilon_{\alpha}-\epsilon_{1}}\right)^2}+\sum _{\alpha=0,L} e^{-i \lambda^{(\alpha)} t} \frac{1}{1+\frac{g^2L}{(\lambda^{(\alpha)}-\epsilon_1)(\lambda^{(\alpha)}-\epsilon_L)}},
\label{photphot}
\end{eqnarray}
where we separated contributions from dark and bright states. In the regime of weak disorder, $\Omega \ll g\sqrt{L}$, a maximum of the absolute value of the first term as a function of $t$ scales as $\sim \Omega^2/g^2L$, while the second term starts to represent collective Rabi oscillations of the spin ensemble as a whole with frequency $g\sqrt{L}$ and the amplitude approaching 1. In the intermediate regime $ g\sqrt{L} \sim \Omega$, amplitudes of oscillations due to these two contributions are generally of the same order.

As an example, Fig. \ref{photon} shows the evolution of the occupation of a single photon state for $L=4$ (a), $L=12$ (b), $L=20$ (c) and at $g=0.05\omega$, $\Omega = 0.1\omega$ calculated numerically. This figure evidences that the dynamics for small spin number is essentially chaotic. However, it becomes more ordered as number of spins grows and the system enters an intermediate regime $\Omega \sim g\sqrt{L}$. Initially, photon is mainly absorbed by a set of dark states, which transform the excitation into the spin subsystem. The photon state occupation in this limit is represented by a superposition of small-amplitude Rabi-like oscillations due to two bright states and quasi-periodical sharp revivals of a period $\sim 1/d$ due to the set of dark states, which release the excitation from the spin subsystem and then absorb it again. These revivals are only partial. The time delay for the first revival increases with the increase of the number of spins in the ensemble due to the growth of the dark state number. The amplitude of oscillations due to bright states grows as $L$ increases, so that in the regime $g\sqrt{L} \gg \Omega$ they have to be transformed to Rabi oscillations of the whole ensemble with frequency $g\sqrt{L}$, while the role of dark states in the dynamics of photon subsystem and for the initial condition considered becomes negligible. Let us stress that, in contrary, they play a dominant role in the dynamics of spin subsystem in this limit for the initial condition considered in the preceding Subsection.

As for the dynamics of spin subsystem, we obtain for the amplitude of probability to have a single spin with energy $\epsilon_m$ excited
\begin{eqnarray}
\langle 0, \downarrow \ldots \downarrow | \sigma_m |\psi(t) \rangle = \sum _{\alpha=0}^{L} e^{-i \lambda^{(\alpha)} t} \frac{1}{1+\sum_{j=1}^{L}\frac{g^2}{(\lambda^{(\alpha)}-\epsilon_j)^2}} \frac{g}{\lambda^{(\alpha)}-\epsilon_m} \notag \\
\simeq \frac{4 d}{\pi^2 g} e^{-i \epsilon_m t} \sum_{n=0}^{\infty}\frac{\sin\left[(2n+1)td/2\right]}{2n+1}+\sum _{\alpha=0,L} e^{-i \lambda^{(\alpha)} t} \frac{1}{1+\frac{g^2L}{(\lambda^{(\alpha)}-\epsilon_1)(\lambda^{(\alpha)}-\epsilon_L)}}\frac{g}{\lambda^{(\alpha)}-\epsilon_m},
\label{photspin}
\end{eqnarray}
where we again separated contributions from dark and two bright states. This probability is generally small.

\subsection{Dynamics of the system with Bell entangled states encoded in the initial moment into a continuum of spins}

Let us now consider the dynamics of the system with constant density of states, when the initial state is one of the Bell states $|\chi_{\pm}\rangle=\frac{1}{\sqrt{2}}(\sigma_{A}^+ \pm \sigma_{B}^+)|\downarrow \downarrow \ldots \downarrow, 0 \rangle$. These states thus can be symmetric and antisymmetric. They represent more sophisticated states which are characterized by a finite quantum entanglement and include more than one spin. The evolution, in this case, is described by Eq. (\ref{explic3}). It is known that, in the case of homogeneous model, the antisymmetric Bell state does not decay at all for any $L$ \cite{Cummings}. It is of interest to explore an influence of broadening on its stability in the case of a mesoscopic ensemble.

\begin{figure}[h]
\includegraphics[width=0.45\linewidth]{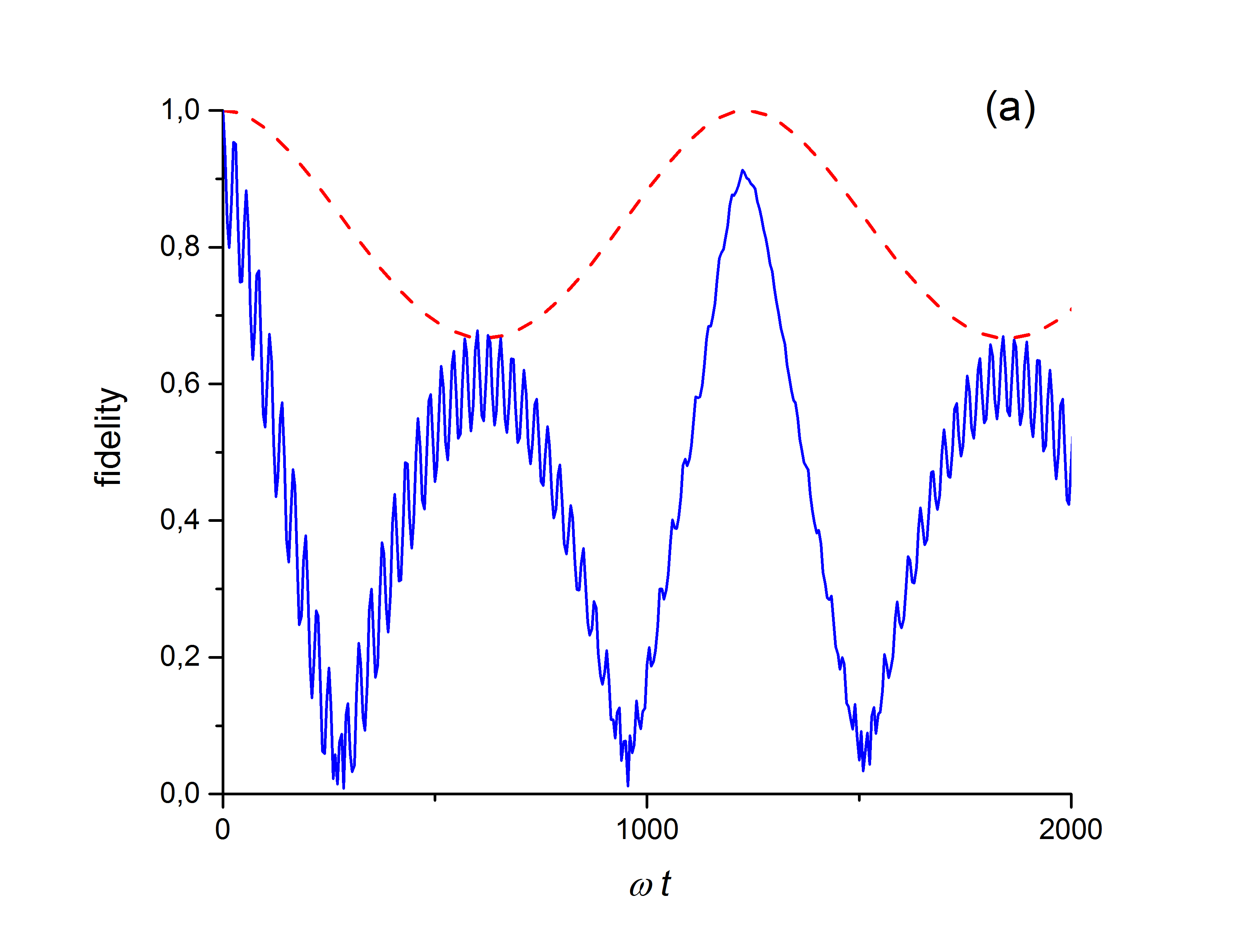}
\includegraphics[width=0.45\linewidth]{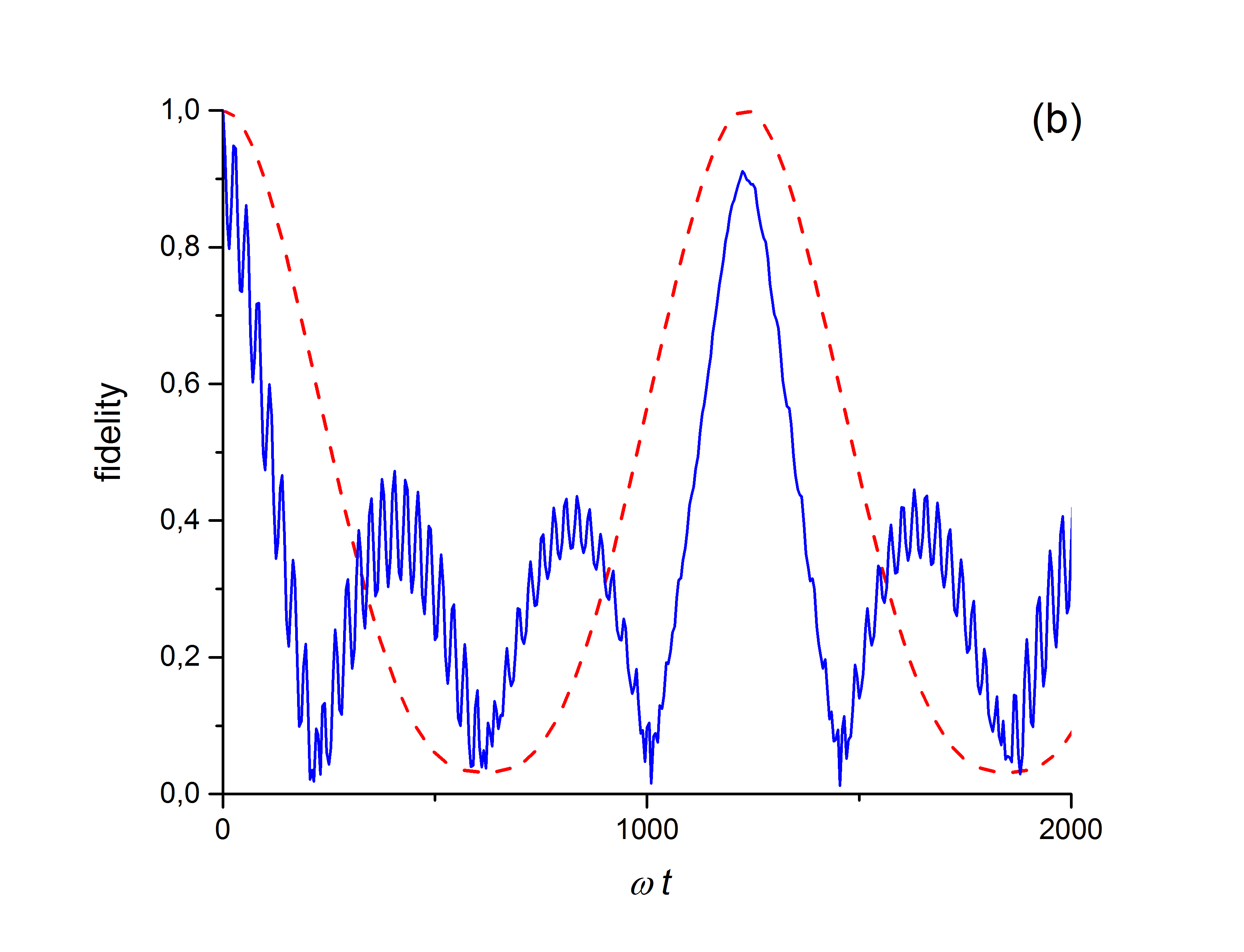}
\includegraphics[width=0.45\linewidth]{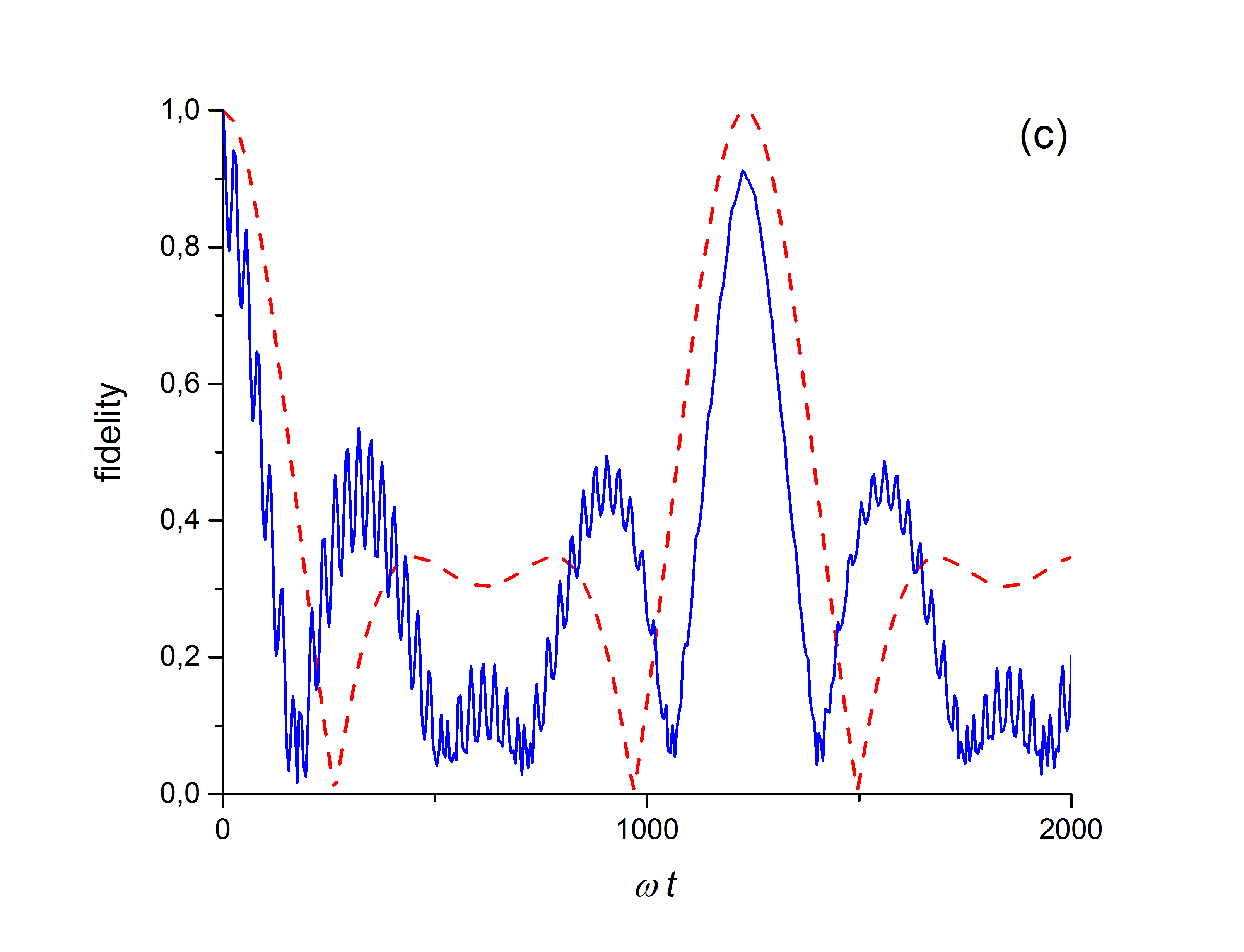}
\caption{
\label{bell}
Evolution of  $|\langle \chi_{\pm} |\psi(t) \rangle|$ for $L=20$, $g=0.05\omega$, $\Omega = 0.1\omega$ and $|\epsilon_B-\epsilon_A|=d$ (a), $2d$ (b), $3d$ (c). Blue solid (red dashed) lines correspond to $\chi_{+}$ ($\chi_{-}$).}
\end{figure}

An important quantity in this context is an overlap between the initial state of the system $\chi_{\pm}$ and its state $|\psi(t) \rangle$ after the beginning of the evolution
\begin{eqnarray}
\langle \chi_{\pm} |\psi(t) \rangle=\frac{1}{2}\sum _{\alpha=0}^{L} e^{-i \lambda^{(\alpha)} t} \frac{1}{1+\sum_{j=1}^{L}\frac{g^2}{(\lambda^{(\alpha)}-\epsilon_j)^2}}
\left[\frac{g}{\lambda^{(\alpha)}-\epsilon_A} \pm \frac{g}{\lambda^{(\alpha)}-\epsilon_B} \right]^2.
\label{fidelity}
\end{eqnarray}
In the large-$L$ limit, this expression reduces to
\begin{eqnarray}
& \langle \chi_{\pm} |\psi(t) \rangle \simeq
\frac{8}{\pi^2}e^{-i (\epsilon_A+\epsilon_B) t/2}
\left(\cos\left(\frac{\epsilon_B-\epsilon_A}{2}t\right) \sum_{n=0}^{\infty}\frac{\cos\left[(2n+1)td/2\right]}{(2n+1)^2}
\pm \frac{d}{(\epsilon_A-\epsilon_B)}\sin\left(\frac{\epsilon_B-\epsilon_A}{2}t\right) \sum_{n=0}^{\infty}\frac{\sin\left[(2n+1)td/2\right]}{2n+1}\right).
\label{fidelities}
\end{eqnarray}
The correspondent evolution of fidelity defined as $|\langle \chi_{\pm} |\psi(t) \rangle|$ and calculated numerically at $L=20$ for three different values of $|\epsilon_B-\epsilon_A|$, $\epsilon_A$ being in a resonance with $\omega$, is plotted in Fig. \ref{bell}. The agreement between the numerics and explicit result (\ref{fidelities}) (not plotted in Fig. \ref{bell}) is again spectacular.

We see that instead of the periodic triangle wave fidelity follows much more complicated evolution. The life time of the state $\chi_{+}$, i.e., the time needed for the fidelity to drop from maximum to the first zero, is of the same order as for the single excited spin, i.e., it is enhanced due to Zeno-like effect. The situation, however, is very different for $\chi_{-}$. Namely, for the minimum possible separation between energies of two spins A and B, $|\epsilon_B-\epsilon_A|=d$, the life time becomes infinite, since the fidelity oscillates periodically but always remains high. Such a behavior in the case of $\chi_{-}$ state is predictable for two isolated spins, but we see that it survives even in presence of an environment of other spins with excitation frequencies close to the resonance, which are expected to strongly interact with a couple of two given spins. In reality, the interaction is strongly suppressed due to a specific structure of a two-spin wave function. The fidelity also does not vanish in the case $|\epsilon_B-\epsilon_A|=2d$, but becomes suppressed in its minimum due to the influence of the intermediate spin. Starting from $|\epsilon_B-\epsilon_A|=3d$, the minimum fidelity reaches zero, but the life time of $\chi_{-}$ state nevertheless remains longer than the lifetime for $\chi_{+}$ state. Thus, $\chi_{-}$ turns out to be much more robust than $\chi_{+}$ especially for small separations between $\epsilon_A$ and $\epsilon_B$. However, the effect of difference between $\epsilon_A$ and $\epsilon_B$ is quite strong and it reflects significant deviations of behavior from the case of homogeneous model. Nevertheless, the evolution of fidelity remains highly sensitive to the plus or minus sign in the Bell state, i.e., quantum interference effects in mesoscopic regime appear as quite robust with respect to broadening.

These results demonstrate that it is possible to stabilize in the spin subsystem collective quantum states involving few spins and not only single spin states. Surprisingly, they can be even more stable than single spin states and are able to support finite entanglement, in principle, for an arbitrary time, which is rather unexpectable, because entanglement is usually considered as a rather fragile entity.

Note that it also follows from Eq. (\ref{fidelity}) that the first-order correction produced by two separated roots is suppressed in the case of $\chi_{-}$ state due to the minus sign in the sum. This is consistent with the results shown in Fig. \ref{bell}.

Next, let us turn to photon subsystem and evaluate the amplitude of probability of having a single photon state. We readily obtain
\begin{eqnarray}
\langle 0, \downarrow \ldots \downarrow | a |\psi(t) \rangle = \frac{1}{\sqrt{2}}\sum _{\alpha=0}^{L} e^{-i \lambda^{(\alpha)} t} \left[\frac{g}{\lambda^{(\alpha)}-\epsilon_A} \pm \frac{g}{\lambda^{(\alpha)}-\epsilon_B} \right] \frac{1}{1+\sum_{j=1}^{L}\frac{g^2}{(\lambda^{(\alpha)}-\epsilon_j)^2}}.
\label{photbell}
\end{eqnarray}
In leading order in $L$, it reduces to
\begin{eqnarray}
\langle 0, \downarrow \ldots \downarrow | a |\psi(t) \rangle \simeq
  \frac{2 \sqrt{2}}{\pi^2} \frac{d}{g} \left(e^{-i \epsilon_A t} \pm e^{-i \epsilon_B t}\right)
\sum_{n=0}^{\infty}\frac{\sin\left[(2n+1)td/2\right]}{2n+1}
\label{photbellfin}
\end{eqnarray}
Due to the presence of the term $\left(e^{-i \epsilon_A t} \pm e^{-i \epsilon_B t}\right)$ the occupation of the single photon state in the case of antisymmetric initial condition and at $\epsilon_A \approx \epsilon_B$ is dramatically reduced compared to the case of a single excited spin considered above, where it is also small. Thus, such a state turns out to be "superdark". This is why $\chi_{-}$ is more robust compared to $\chi_{+}$. In the case of homogeneous model, $\chi_{-}$ becomes a true Hamiltonian eigenstate totally decoupled from light (subradiant mode \cite{Cummings}).

The results of this Subsection might be also used for testing quantum mechanical nature of artificial spin-cavity systems including a manifestation of quantum interference effects, which are crucial for an unambiguous demonstration of "quantumness" of such engineered macroscopic systems. The entangled state between two qubits can be created via additional tunable cavities coupled to these particular qubits. By performing a two-qubit tomography it is then possible to follow the entanglement dynamics and to deeply probe coherent properties of qubits-cavity coupled systems. Particularly, the evolution must be sensitive to the symmetry of Bell state (plus or minus sign).

\section{Effects of statistics}

Let us briefly discuss the effects arising due to deviation from equally-spaced distribution towards Gaussian or similar types of statistics characterized by smooth tails. We found numerically that if the typical energy scale associated with the deviation from equally-spaced distribution (such as mean variance for Gaussian distribution) is much larger than $g$, no qualitative change of behavior for dynamics starting from excitations within spin subsystem is observed at short times $\sim 2\pi d_0$, $d_0$ being an inverse maximum density of states. Namely, leading order contributions to the quantities found above are qualitatively very similar. This happens because most strong interaction between the photon and spin subsystems is due to spins with energies close to the resonance with $\omega$, while the influence of remaining spins become larger at long times. Nevertheless, subdominant contributions do change since no well defined bright state survives under Gaussian-like distributions with tails; this, for instance, results in absence of Rabi-like oscillations, which are transformed into small amplitude chaotic dependencies similar to the ones visible in Fig. \ref{rand}.

However, rather significant changes do occur provided the scale associated with the deviation from equally-spaced distribution is lowered to $\lesssim g$. In this case, dynamics becomes more irregular. Nevertheless, the typical relaxation time of a single spin excited remains to be long. Moreover, our general conclusion about extreme robustness of antisymmetric Bell state $\chi_{-}$, which do not decay at all for neighboring spins $A$ and $B$, remains valid as well.

\section{Conclusions}

We studied dynamics of mesoscopic ensemble of qubits coupled to a single mode cavity and concentrated on effects of disorder in qubits excitation frequencies and the regime of a weak excitation. In particular, we analyzed how collective properties of such a coupled system are formed as the number of qubits in the ensemble grows. This is done using a Bethe ansatz solution of Dicke model, which provides a direct access to Hamiltonian exact eigenstates as well as simple and pictorial understanding based on them.

We found that dark states weakly coupled to light gradually emerge and start to play a very important role upon the crossover from few-qubit systems to large ensembles. They are similar to subradiant modes of homogeneous Dicke model, which are totally decoupled from the light. The role of dark states in the free evolution dynamics is more important for initial conditions corresponding to excitations within qubit subsystem.

Our main conclusions are as follows:

(a) Despite of inhomogeneous broadening, single qubit excitation becomes stabilized in the infinite qubit number limit through what can be referred to as Zeno-like effect. It is induced by the gradual formation of dark states.

(b) Surprisingly, antisymmetric entangled Bell state constructed from the individual states of two qubits can be even more stable than single qubit excitation, provided excitation frequencies of these two qubits are not far from each other in the energy space. However, as separation grows, inhomogeneous broadening is able to suppress this effect. Nevertheless, the evolution remains  highly sensitive to the symmetry of the total wave function, which highlights a nontrivial role of quantum interference effects in mesoscopic regime.

(c) The effect of finite number of spins is generally manifested through partial revivals of the initial state, the period of revivals being proportional to the number of qubits. This scenario explains how Rabi-like oscillations in the limit of just few qubits are transformed into collective and highly cooperative behavior in the limit of many qubits.

(d) The role of dark states is less essential for initial condition corresponding to the single photon. They however are still of importance in the mesoscopic regime, when they are able to absorb photon into the qubit subsystem and then to release it quasiperiodically in time. A characteristic period of such revivals grows with the number of qubits, while the amplitude lowers. They coexist with Rabi-like oscillations between the whole ensemble of qubits and photon field so that in the infinite number limit such oscillations are naturally recovered.

Our results provide additional insights to physics of qubits-cavity coupled systems and might be of importance for quantum states engineering, protection, and storage. In particular, we believe that our theoretical predictions can be used to deeply probe both coherent and collective properties of mesoscopic ensembles of artificial spins coupled to cavities. For instance, Zeno-like effect is definitely linked to the formation of collective properties of the ensemble despite of the disorder in excitation frequencies. On the other hand, a sensitivity of the evolution of the entangled Bell state to the symmetry of the wave function is based on quantum interference effects, while experimental demonstration of such effects is crucial for the unambiguous evidence of true "quantumness" of engineered macroscopic artificial systems.

\begin{acknowledgments}
 Useful comments by A. N. Rubtsov, A. A. Elistratov, and S. V. Remizov are acknowledged. This work is supported by Advanced Research Foundation (project no. 7/076/2016-2020). W. V. P. acknowledges a support from RFBR (project no. 15-02-02128). Yu. E. L. acknowledges a support from RFBR (project no. 17-02-01134).
\end{acknowledgments}

\appendix

\section{Time dependence}

Let us consider the dynamics of the system starting from some initial state $|\psi(t=0)\rangle$ corresponding to $M=1$ this number being conserved during the evolution. We expand the time-dependent wave function over the orthonormal basis $|\varphi_{\alpha} \rangle$ as
\begin{eqnarray}
|\psi(t) \rangle=\sum _{\alpha=0}^{L} C_{\alpha}(t) |\varphi_{\alpha} \rangle,
\label{initexp}
\end{eqnarray}
where
\begin{eqnarray}
C_{\alpha}(t=0)=\langle \varphi_{\alpha} | \psi(t=0) \rangle.
\label{initoverlap}
\end{eqnarray}
These coefficients at arbitrary $t>0$ can be readily found from the Schr\"{o}dinger equation
\begin{eqnarray}
C_{\alpha}(t)=C_{\alpha}(t=0) \exp (-i \lambda^{(\alpha)} t).
\label{initoverlap11}
\end{eqnarray}

Below we consider several initial conditions and general distribution of spin energies. These general results are then used to analyze system's dynamics explicitly for an equally-spaced distribution of spin energies.

\subsection{Single spin excited}

Let us obtain the time dependent wave function corresponding to the initial state defined as
\begin{eqnarray}
|\psi(t=0)\rangle= \sigma_{A}^+|\downarrow \downarrow \ldots \downarrow, 0 \rangle,
\label{init}
\end{eqnarray}
which describes a single spin excited, while all the remaining spins are in their ground states and there is no photon in the system. Such states have a direct relation to states engineered in Ref. by a spectral hole burning technique. It is of interest to explore its dynamics as the number of spins in the system grows. Note that in the limit $L=1$ there appear Rabi oscillations between the single spin and photon field. The naive expectation is that an addition of extra spins would just lead to a sort of a chaotization of Rabi oscillations. We will show that this is not the case.

We readily obtain from Eq. (\ref{initoverlap})
\begin{eqnarray}
C_{\alpha}(t=0)=\langle \varphi_{\alpha} | \psi(t=0) \rangle = \frac{1}{\sqrt{\langle \Phi_{\alpha}|\Phi_{\alpha} \rangle}}\frac{g}{\lambda^{(\alpha)}-\epsilon_A}.
\label{initoverlap}
\end{eqnarray}
In the explicit form, time dependent wave function reads
\begin{eqnarray}
|\psi(t) \rangle=\sum _{\alpha=0}^{L} e^{-i \lambda^{(\alpha)} t} \frac{g}{\lambda^{(\alpha)}-\epsilon_A} \frac{1}{1+\sum_{j=1}^{L}\frac{g^2}{(\lambda^{(\alpha)}-\epsilon_j)^2}}(a^{\dagger }+\sum_{j=1}^{L}\frac{g}{\lambda^{(\alpha)}-\epsilon_j}\sigma_{j}^+)|\downarrow \downarrow \ldots \downarrow, 0 \rangle.
\label{explic}
\end{eqnarray}

\subsection{Single photon}

We may also consider the initial condition of another type, which corresponds to the single photon in a cavity
\begin{eqnarray}
|\psi(t=0)\rangle = a^{\dagger }|\downarrow \downarrow \ldots \downarrow, 0 \rangle.
\label{initphot}
\end{eqnarray}
From Eq. (\ref{initoverlap}), we again find
\begin{eqnarray}
C_{\alpha}(t=0)=\langle \varphi_{\alpha} | \psi(t=0) \rangle = 1.
\label{initoverlapphot}
\end{eqnarray}
In the explicit form, time dependent wave function is
\begin{eqnarray}
|\psi(t) \rangle=\sum _{\alpha=0}^{L} e^{-i \lambda^{(\alpha)} t}  \frac{1}{1+\sum_{j=1}^{L}\frac{g^2}{(\lambda^{(\alpha)}-\epsilon_j)^2}}(a^{\dagger }+\sum_{j=1}^{L}\frac{g}{\lambda^{(\alpha)}-\epsilon_j}\sigma_{j}^+)|\downarrow \downarrow \ldots \downarrow, 0 \rangle.
\label{explicphot}
\end{eqnarray}

\subsection{Bell state encoded into the spin subsystem}

Let us now consider a dynamics of the system with the initial state being one of the two Bell states encoded into two spins within the spin subsystem
\begin{eqnarray}
|\chi_{\pm}\rangle=\frac{1}{\sqrt{2}}(\sigma_{A}^+ \pm \sigma_{B}^+)|\downarrow \downarrow \ldots \downarrow, 0 \rangle.
\label{kai}
\end{eqnarray}
Using the developed approach, we arrive at the explicit form of the time dependent wave function
\begin{eqnarray}
|\psi(t) \rangle=\frac{1}{\sqrt{2}}\sum _{\alpha=0}^{L} e^{-i \lambda^{(\alpha)} t} \left[\frac{g}{\lambda^{(\alpha)}-\epsilon_A} \pm \frac{g}{\lambda^{(\alpha)}-\epsilon_B} \right] \frac{1}{1+\sum_{j=1}^{L}\frac{g^2}{(\lambda^{(\alpha)}-\epsilon_j)^2}}(a^{\dagger }+\sum_{j=1}^{L}\frac{g}{\lambda^{(\alpha)}-\epsilon_j}\sigma_{j}^+)|\downarrow \downarrow \ldots \downarrow, 0 \rangle.
\label{explic3}
\end{eqnarray}

\section{Roots for equally-spaced distribution}

For the confined roots, it is convenient to represent $\lambda^{(\alpha)}$ as $\epsilon_{\alpha}+\delta_{\alpha}$, where $\delta_{\alpha} < d$. In order to find $\delta_{\alpha}$, we split the sum in Eq. (\ref{Bethesingle}) into two contributions
\begin{eqnarray}
\frac{\lambda^{(\alpha)}-\omega}{g^2} = \sum_{k=0}^{\min(\alpha-2,L-\alpha)} \left[ \frac{1}{dk+d+\delta_{\alpha}}-\frac{1}{dk-\delta_{\alpha}}\right] + \notag \\
\left[\sum_{k=\min(\alpha-2,L-\alpha)+1}^{\alpha-2}  \frac{1}{dk+d+\delta_{\alpha}}
- \sum_{k=\min(\alpha-2,L-\alpha)+1}^{L-\alpha}  \frac{1}{dk-\delta_{\alpha}}\right].
\label{Betheeqspaced}
\end{eqnarray}
The first contribution is over the energies $\epsilon_j$ extending symmetrically from $\epsilon_{\alpha}$ in both sides, while the second one includes contributions of the remaining energies $\epsilon_j$. Since we consider the regime of large $L \gg 1$, we may replace upper limits of summation in the first contribution by $+\infty$. This term then reduces to
\begin{eqnarray}
\sum_{k=0}^{+\infty} \left[ \frac{1}{dk+d+\delta_{\alpha}}-\frac{1}{dk-\delta_{\alpha}}\right] = \frac{\pi}{d} \cot\left(\pi\frac{\delta_{\alpha}}{d}\right).
\label{cot}
\end{eqnarray}
We may also replace summation by integration in the remaining terms in Eq. (\ref{Betheeqspaced}) as well as $\lambda^{(\alpha)}$ by $\epsilon_{\alpha}$ in its left-hand side. We then arrive at the expression for $\delta_{\alpha}$ given by
\begin{eqnarray}
\delta_{\alpha} \simeq \frac{d}{\pi}  \cot^{-1} \left( \frac{1}{\pi} \left[\frac{d}{g^2}(\epsilon_{\alpha} - \omega) + \ln \frac{\epsilon_L-\epsilon_{\alpha}}{\epsilon_{\alpha}-\epsilon_{1}} \right] \right).
\label{cot}
\end{eqnarray}
An important special case is a distribution centered around $\omega$, which leads to $\delta_{\alpha} \simeq d/2$ for $\epsilon_{\alpha}$ approaching $\omega$.

Using a similar method, we evaluate $\langle \Phi_{\alpha}|\Phi_{\alpha} \rangle$ as
\begin{eqnarray}
\langle \Phi_{\alpha}|\Phi_{\alpha} \rangle \simeq g^2 \frac{\pi^2}{d^2} \frac{1}{\sin^2 \left(\frac{\pi \delta^{(\alpha)}}{d}\right)}.
\label{norm1}
\end{eqnarray}
The expression of $\sin^2 (\frac{\pi \delta^{(\alpha)}}{d})$ can be readily obtained from Eq. (\ref{cot}) yielding
\begin{eqnarray}
\langle \Phi_{\alpha}|\Phi_{\alpha} \rangle \simeq g^2 \frac{\pi^2}{d^2} \left[1+\frac{1}{\pi^2}\left(\ln\frac{\epsilon_L-\epsilon_{\alpha}}{\epsilon_{\alpha}-\epsilon_{1}}\right)^2\right].
\label{norm2}
\end{eqnarray}

Let us now consider two additional roots of Eq. (\ref{Bethesingle}), which we denote as $\lambda^{(0)}$ and $\lambda^{(L)}$, while $\lambda^{(0)} < \epsilon_1 $ and $\lambda^{(L)} > \epsilon_L $. In order to find these roots, we may replace summation by integration in this equation, which is allowed provided $\epsilon_1-\lambda^{(0)} \gg d$ and $\lambda^{(L)}-\epsilon_L \gg d$. The equations for $\lambda^{(0)}$ and $\lambda^{(L)}$ are identical and they read as
\begin{eqnarray}
\lambda^{(0,L)}-\omega \simeq \frac{g^2L}{\Omega} \ln\frac{\lambda^{(0,L)}-\epsilon_1}{\lambda^{(0,L)}-\epsilon_L}.
\label{transc}
\end{eqnarray}
These are transcendental equations, which can be solved numerically. Let us stress that the dependence of $\lambda^{(0,L)}$ on $g$ is nonanalytic. The norms for two corresponding eigenfunctions are
\begin{eqnarray}
\langle \Phi_{0,L}|\Phi_{0,L} \rangle \simeq 1+ \frac{g^2 L}{(\lambda^{(0,L)}-\epsilon_1)(\lambda^{(0,L)}-\epsilon_L)}.
\label{normsepar}
\end{eqnarray}


\end{document}